\documentclass{llncs} 

\usepackage{amssymb}
\usepackage{textcomp}
\usepackage{comment}
\usepackage{color}
\usepackage{graphicx}
\usepackage{euscript}
\usepackage{ dsfont }
\usepackage{amsmath}
\usepackage{booktabs}
\usepackage{multirow}
\usepackage[table,xcdraw]{xcolor}
\newcommand{\Beginproof}{{\em Proof.}  }
\newcommand{\Endproof}{\hfill$\Box$\\}

\begin{document}

\title{Quantum Online Algorithms with Respect to Space Complexity}
\author{Kamil~Khadiev$^{1,2}$\and Aliya Khadieva$^2$\and Ilnaz Mannapov$^2$}

\institute{ University of Latvia, Riga, Latvia      
       \and       
Kazan Federal University, Kazan, Russia      
                \\ \email{kamilhadi@gmail.com, aliya.khadi@gmail.com, ilnaztatar5@gmail.com} 
}

 \maketitle

\begin{abstract}
Online algorithm is a well-known computational model. We introduce quantum online algorithms and investigate them with respect to a competitive ratio in two points of view: space complexity and advice complexity. We start with exploring a model with restricted memory and show that quantum online algorithms can be better than classical ones (deterministic or randomized) for sublogarithmic space (memory), and they can be better than deterministic online algorithms without restriction for memory.  Additionally, we consider polylogarithmic space case and show that in this case, quantum online algorithms can be better than deterministic ones as well.
 
Another point of view to the online algorithms model is advice complexity.  So, we introduce quantum online algorithms with a quantum channel with an adviser. Firstly, we show that quantum algorithms have at least the same computational power as classical ones have. And we give some examples of quantum online algorithms with advice. Secondly, we show that if we allow to use shared entangled qubits (EPR-pairs), then quantum online algorithm can use two times less advise qubits comparing to a classical one. We apply this approach to the well-known Paging Problem. 

\end{abstract}

\section{Introduction}
Online algorithms are a well-known computational model for solving optimization problems. The peculiarity is that algorithm reads an input piece by piece and should return an answer piece by piece immediately, even if an answer can depend on feature pieces of the input. An online algorithm should return an output for minimizing an objective function. There are different methods to define the effectiveness of the algorithms \cite{bil2009,bil2015,dl2005}. But a most standard is the competitive ratio \cite{kmrs86,st85}. It is a ratio between output's price for an online algorithm and optimal offline algorithms.
 
We suggest a new model for online algorithms, {\em quantum online algorithms} that use a power of quantum computing for solving online minimization problem. In this model, an {\em algorithm} can have quantum and classical part. Additionally, we consider restricted version of the algorithms, {\em pure quantum online algorithms}. These algorithms have only quantum part.
We focus on two measures of complexity: space complexity and advice complexity.

 When we consider space, we discuss the online algorithms with restricted memory. For this kind of algorithms, we allow to use only $s$ bits of memory, for given integer $s$. Another point of view to the same model is streaming algorithms for an online minimization problem. And such classical models were considered in  \cite{bk2009,gk2015,blm2015}. In this case, we show that quantum online algorithms with single qubit can be better than any classical (deterministic or randomized) online algorithms with sublogarithmic memory.  And this quantum algorithm still can be better than any deterministic online algorithms without a restriction for memory. 
It is also interesting to investigate the model with polylogarithmic memory or logarithmic memory (LOGSPACE). Here the algorithm can use  $(\log n)^{O(1)}$ bits of memory. We show that for $(n,k,r,w)$-Parity for Number of Equality Hats problem the quantum algorithm can be better than any deterministic one.
Additionally, we show that quantum online algorithm can simulate a classical (deterministic or randomize) one with almost same memory. It means that quantum model has at least the same computational power. Note that for other models, quantum and classical cases can be incomparable, for example, communication complexity model \cite{k2000} or Ordered Binary Decision Diagrams \cite{ss2005}. 
  Online algorithms with restricted memory are similar to streaming algorithms \cite{l2006,gkkrw2007}, Branching programs \cite{Weg00}, Automata \cite{AY12,AY15}. Researchers also compare classical and quantum cases for these models  
\cite{agkmp2005,agky14,agky16,g15,ss2005,kk2017,AY12,AY15,l2006,aakk2017}, \cite{ikpvy2017,gkkrw2007}.  
 
 Another interesting complexity measure for the model with respect to the competitive ratio is {\em advice complexity} \cite{k2016,bfklm2016,bfklm2017}. In this model online algorithm gets some bits of advice about an input. Trusted {\em Adviser} sending these bits knows the whole input and has an unlimited computational power. The question is ``how many advice bits are enough to reduce competitive ratio or to make the online algorithm as the same effective as the offline algorithm in the worst case?''. This question has different interpretations. One of them is ``How many information an algorithm should know about a future for solving a problem effectively?''. Another one is ``If we have an expensive channel which can be used for pre-processed information about the future, then how many bits we should send by this channel to solve a problem effectively?''. Researchers pay attention to deterministic and  probabilistic or randomized online algorithms with advice \cite{h2005,kt2006,mr2006,k2016}.

We suggest {\em quantum online algorithms with advice}. In this model, an {\em algorithm} has a quantum channel with {\em Adviser}.
  We consider two kinds of the model. The first one is an algorithm with private qubits. In this case, an {\em algorithm} and an {\em Adviser} have not shared qubits. Advice bits are only communication between these two ``players''. For this model, we show that we also can simulate a classical (deterministic or randomized) online algorithm with the same number of advice bits. And we demonstrate examples of quantum online algorithms with advice for special $(n,k,r,w)$-Parity for Number of Hats problem.
 The second kind of algorithms is a model with shared entangled qubits or EPR-pairs. Before the computational process, we prepare pairs of entangled qubits (EPR-pairs), and for each pair, we give one qubit to the {\em algorithm} and another to the {\em Adviser}. Then the {\em Adviser} sends twice less advice qubits using a trick from \cite{bw92}. We apply this idea to the well-known {\em Paging problem} \cite{h2005,be2005}.
 
  The paper is organized in the following way. We present definitions in Section \ref{sec:prlmrs}.  In Section \ref{sec:space} we explore online algorithms with restricted memory. The quantum online algorithms with advice are explored in Section \ref{sec:advice}. We focus on a model with private quantum bits in Section \ref{sec:privat} and with shared entangled qubits in Section \ref{sec:shared}.  

\section{Preliminaries}\label{sec:prlmrs}
Firstly, let us define an online optimization problem. All following definitions we give with respect to \cite{k2016}.
\begin{definition}[Online Minimization Problem] An online minimization problem consists of a set $\cal{I}$ of inputs and a cost function. Every input $I\in\cal{I}$ is a sequence of requests $I = (x_1, \dots , x_n)$. Furthermore, a set of feasible outputs (or solutions) is associated with every $I$; every output is a sequence of answers $O = (y_1, \dots , y_n)$. The cost function assigns a positive real value $cost(I, O)$ to every input $I$ and any feasible output $O$. For every input $I$, we call any feasible output $O$ for $I$ that has the smallest possible cost (i. e., that minimizes the cost function) an optimal solution for $I$. \end{definition}  

Let us define an online algorithm for this problem as an algorithm which gets requests $I$ one by one and should return answers $O$ immediately, even if an optimal solution can depend on future requests.
\begin{definition}[Deterministic online algorithm] Consider an input $I$ of an online minimization problem. An online algorithm $A$ computes the output sequence $A(I) = (y_1,\dots , y_n)$ such that $y_i$ is computed from $x_1, \dots , x_i$, $y_1, \dots , y_{i-1}$. We denote the cost of the computed output by $cost(A(I))= cost(I,A(I))$. \end{definition}  

  This setting can also be regarded as a request-answer game: an adversary generates requests, and an algorithm has to serve them one at a time \cite{a1996}.  
   
  As the main measure of quality of an online algorithm, we use a competitive ratio. It is the ratio of two costs: cost for an online algorithm's solution; and cost for an optimal offline algorithm solution. We consider the worst case.  

\begin{definition}[Competitive Ratio]
An online algorithm $A$ is $c$-competitive if there exists a non-negative constant $\alpha$ such that, for every input $I$, we have: $cost(A(I)) \leq c \cdot cost(Opt(I)) + \alpha,$ where $Opt$ is an optimal offline algorithm for the problem. We also call $c$ the competitive ratio of $A$. If $\alpha = 0$, then $A$ is called strictly $c$-competitive; $A$ is optimal if it is strictly $1$-competitive. \end{definition}  

  Let us define an Online Algorithm with Advice. We can say, that advice is some information about the future input.
    
  \begin{definition}[Online Algorithm with Advice] Consider an input $I$ of an online minimization problem. An online algorithm $A$ with advice computes the output sequence $A^{\phi}(I) = (y_1, \dots , y_n)$ such that $y_i$ is computed from $\phi, x_1, \dots , x_i$, where $\phi$ is the message from Adviser, who knows the whole input.
  $A$ is $c$-competitive with advice complexity $b=b(n)$ if there exists a non-negative constant $\alpha$ such that, for every $n$ and for any input $I$ of length at most $n$, there exists some $\phi$ such that $cost(A^{\phi}(I)) \leq c \cdot cost(Opt(I)) + \alpha$ and length of $\phi$ is at most $b$ bits. \end{definition}  

  Next, let us define a randomized online algorithm.  
\begin{definition}[Randomized Online Algorithm] Consider an input $I$ of an online
minimization problem. A randomized online algorithm $R$ computes the output sequence
$R^{\psi} := R^{\psi}(I) = (y_1,\cdots, y_n)$ such that $ y_i$ is computed from $\psi, x_1, \cdots, x_i$, where $\psi$ is the content of a random tape, i. e., an infinite binary sequence, where every bit is chosen uniformly at random and independently of all the others. By $cost(R^{\psi}(I))$ we denote the random variable expressing the cost of the solution computed by $R$ on $I$.
$R$ is $c$-competitive in expectation if there exists a non-negative constant $\alpha$ such that, for every $I$, $\mathds{E}[cost(R^{\psi}(I))] \leq c \cdot cost(Opt(I)) + \alpha,$
where $Opt$ is an optimal offline algorithm for the problem.
\end{definition}
We consider online algorithms that base their computation on both advice bits and randomness.

\begin{definition}[Randomized Online Algorithm with Advice] 
Consider an input $I$ of an online
minimization problem. A randomized online algorithm $R$ computes the output sequence
$R^{\psi,\phi} := R^{\psi, \phi}(I) = (y_1,\cdots, y_n)$ such that $ y_i$ is computed from $\psi, \phi, x_1, \cdots, x_i$, where $\psi$ is the content of a random tape, i. e., an infinite binary sequence, where every bit is chosen uniformly at random and independently of all the others, and $\phi$ is the message from Adviser, who knows whole input. By $cost(R^{\psi,\phi}(I))$ we denote the random variable expressing the cost of the solution computed by $R$ on $I$.
$R$ is $c$-competitive in expectation if there exists a non-negative constant $\alpha$ such that, for every $I$ of length at most $n$, there is $\phi$ such that, $\mathds{E}[cost(R^{\psi,\phi}(I))] \leq c \cdot cost(Opt(I)) + \alpha,$
where $Opt$ is an optimal offline algorithm for the problem and length of $\phi$ is at most $b$ qubits. 
\end{definition}

A randomized online algorithm with advice is allowed to make random choices (i.e., ``toss coins'') to determine its actions and the advice scheme. Formally, then, a randomized online algorithm with advice is a probability distribution over deterministic online algorithms with advice.

Now we are ready to define a {\em quantum online algorithm}. You can read more about quantum computation in \cite{AY15}

\begin{definition}[Quantum Online Algorithm] Consider an input $I$ of an online
minimization problem. A quantum online algorithm $Q$ computes the output sequence
$Q(I) = (y_1,\cdots, y_n)$ such that $ y_i$ is computed from $x_1, \cdots, x_i$. $Q$ can have classical and quantum part. The algorithm can measure qubits several times during computation.  If the algorithm has not classical part, then we call it ``pure quantum online algorithm''.  By $cost(Q(I))$ we denote the cost of the solution computed by $Q$ on $I$. Note that quantum computation is probabilistic process.
$Q$ is $c$-competitive in expectation if there exists a non-negative constant $\alpha$ such that, for every $I$, $\mathds{E}[cost(Q(I))] \leq c \cdot cost(Opt(I)) + \alpha,$
where $Opt$ is an optimal offline algorithm for the problem.
\end{definition}

Let us define a {\em quantum online algorithm with advice}. We define two kinds of the model: a model with shared and a model with private qubits. (We allow or do not allow to share qubits between {\em Adviser} and {\em Algorithm} before the start of computation.)  

\begin{definition}[Quantum Online Algorithm with Advice and Private qubits] Consider an input $I$ of an online
minimization problem. A quantum online algorithm $A$ with advice and private qubits (or quantum online algorithm with advice) computes the output sequence $A^{|\phi\rangle}(I) = (y_1, \dots , y_n)$ such that $y_i$ is computed from $|\phi\rangle, x_1, \dots , x_i$, where $|\phi\rangle$ is the quantum message from Adviser, who knows the whole input. $A$ can have classical and quantum part and do not have any restrictions on measurement.
$A$ is $c$-competitive with advice complexity $b=b(n)$ if there exists a non-negative constant $\alpha$ such that, for every $n$ and for any input $I$ of length at most $n$, there exists some $|\phi\rangle$ such that $\mathds{E}[cost(A^{|\phi\rangle}(I))] \leq c \cdot cost(Opt(I)) + \alpha$ where, as above, $Opt$ is an optimal offline algorithm for the problem and length of $|\phi\rangle$ is at most $b$ qubits.\end{definition}  
  \begin{definition}[Quantum Online Algorithm with Advice and Shared qubits] Consider an input $I$ of an online minimization problem. A quantum online algorithm $A$ with advice, and shared qubits is a quantum online algorithm with advice, but before the process of computation, the Algorithm prepares shared qubits (may be entangled) and sends part of them to the Adviser.\end{definition}

Let us define online algorithms with restricted memory. 
Let deterministic online algorithm $A_s$ be an algorithm which uses at most $s$ bits of memory on processing any input $I$. We can define similar restrictions for randomized algorithms and algorithms with advice.  
Let quantum online algorithm $Q_{s,t}$ be an algorithm which uses at most $s$ classical bits of memory and $t$ quantum bits of memory on processing any input $I$. Let pure quantum online algorithm $Q_{t}$ be an algorithm using at most $t$ quantum bits of memory on processing any input $I$.    

\section{Space Complexity for Quantum Online Algorithms}\label{sec:space}
Let us focus on space complexity of online algorithms. It is interesting to analyze the size of memory that is required by the algorithm. In a case of the restricted memory, a quantum algorithm can be better than classical ones (deterministic or probabilistic). We present this result in Theorems \ref{th:phn-quantum-s}, \ref{th:phn-deterministic-s} and \ref{th:pnh-random-s}. And the quantum algorithm can be still better than any deterministic online algorithm without a restriction to a size of memory (Theorem \ref{th:deterministic-pnh-s}).

Let us consider the special problem which allows us to show the separation: {\em $(n,k,r,w)$-Parity for Number of Hats} ($(n,k,r,w)$-PNH). 

Definition of $(n,k,r,w)$-PNH problem is based on definition of $PartialMOD_n^k$ function from \cite{AY12,agky14,agky16}. Feasible inputs for the problem are $X=(x_1,\dots,x_n)$, for $x_1,\dots,x_{n}\in\{0,1\}$ such that $\#_1(X)=v\cdot 2^k$, where $\#_1(X)$ is the number of $1$s and $v\geq 2$ is a positive integer. $PartialMOD_n^k(X)=v$ $\mod$ $2$.

Firstly, let us describe $(n,k,r,w)$-PNH problem informally.
There are  $3$ guardians and $3$ prisoners. They stay one by one in a line ``$G_1P_1G_2P_2G_3P_3$, $G_i$ is guardian and $P_i$ is prisoner. Prisoner $P_i$ has an input $X_i$ of length $m_i$ and computes function $PartialMOD_{m_i}^k(X_i)$ for $i \in \{1 ,2, 3\}$. If the result is $1$ then he paints his hat in black. Otherwise, he paints it in white. Each guardian wants to know if the number of following black hats is odd or even. The cost of right guardian's answer is $r$, and the cost of the wrong answer is $w$. We want to minimize the cost of output, and assume that $r<w$. 

Formal definition of $(n,k,r,w)$-PNH is following:
Feasible inputs for the problem are $I=(x_1,\dots,x_n)$ of length $n$ such that $n=m_1+m_2+m_3+3$, for some integer $m_1,m_2,m_3\geq 2^{k+1}$. It is guarantied that $I$ is always such that $I=2,X_1,2,X_2,2,X_3$, where $X_i\in\{0,1\}^{m_i}$, for $i\in\{1,2,3\}$. Additionally,  $\#_1(X_i)=v_i\cdot 2^k$, where $v_i$ is some integer, $i\in\{1,2,3\}$. Let $O$ be output of $(n,k,r,w)$-PNH and $O'=y_1,y_3,y_3$ be output bits corresponding to input variables with value $2$ (in other words, variables of guardians).  Output $y_1$ corresponds to $x_1$, $y_2$ corresponds to $x_{2+m_1}$ and , $y_3$ corresponds to $x_{3+m_1+m_2}$. Let 
$z_j(I)=\bigoplus_{i=j}^3 PartialMOD_{m_i}^k(X_i)$. The cost $cost(I,O)=r$, if $y_j=z_j$ for all $j\in\{1,2,3\}$, and $cost(I,O)=w$ otherwise. We consider numbers $r$ and $w$ such that $r<w$. 

Let us present a pure quantum online algorithm which uses single qubit of memory for this problem. It uses ideas from quantum automata \cite{AY12} and branching programs theories \cite{agky14,agky16}.

{\bf Algorithm 1. (Quantum Online Algorithm for $(n,k,r,w)$-PNH)} The pure quantum algorithm $Q_1$ uses single qubit.

{\em Step 1.} The algorithm emulates guessing for $z_1(I)$. $Q_1$ starts on a state $\frac{1}{\sqrt{2}}|0\rangle+\frac{1}{\sqrt{2}}|1\rangle$. And it measures the qubit before reading any input variables. It gets $|0\rangle$ or $|1\rangle$ with equal probability. The result of measurement is $y_1$. 

{\em Step 2.} The algorithm reads $X_1$. Let angle $\alpha=\pi/2^{k+1}$. Then $Q_1$ rotates the qubit by an angle $\alpha$, if the algorithm meets $1$. And it does not do anything otherwise.

{\em Step 3.} If $Q_1$ meets $2$ then it measures the qubit $|\psi\rangle=a|0\rangle+b|1\rangle$. If $PartialMOD_{m_1}^k(X_1)=1$ then the qubit is rotated by an angle $\pi/2+v\cdot \pi$, for some integer $v$, else the qubit is rotated by an angle $w\cdot \pi$, for some integer $w$. If $y_1=1$, then $a\in\{1,-1\}$ and $b=0$. And if $y_1=0$, then $a=0$ and $b\in\{1,-1\}$. The result of measurement is $y_2$. 

{\em Step 4.} The step is similar to Step 2, but algorithm reads $X_2$.

{\em Step 5.} The step is similar to Step 3, but algorithm outputs $y_3$.

{\em Step 6.} The algorithm reads and skips the last part of the input. $Q_1$ does not need these variables, because  it guesses $y_1$ and using this value we already can obtain $y_2$ and $y_3$ without $X_3$.

Assume, that algorithm did right guess that $z_1=y_1$. So, if the parity of the passing part is the same as the parity of the future part of the input, then the algorithm returns the right answer with probability $1$. And if the guess is not correct and $z_1\neq y_1$, then the algorithm returns a wrong answer with probability $1$.  

 With equal probabilities $0.5$ we have  $z_1=y_1$ or  $z_1\neq y_1$. Thus competitive ratio is $(0.5 \cdot r+0.5 \cdot w)/r=(r+w)/(2r)$.
 
As a result we have the following theorem:

\begin{theorem}\label{th:phn-quantum-s} 
There is $(r+w)/(2r)$-competitive in expectation pure quantum algorithms for $(n,k,r,w)$-PNH Problem $Q_1$ with a single qubit of memory. \end{theorem}

At the same time, if a deterministic online algorithm for $(n,k,r,w)$-PNH Problem uses less than $k$ bits; then it is $(w/r)$-competitive. To show this claim, let us discuss properties of $PartialMOD_n^k$ function.

\begin{lemma}\label{lm:pp-deterministic-s}
Let integer $s,k$ be such that $s<k=o(\log(n))$, where $n$ is the length of input. Then there is no deterministic algorithm that reads an input variable by variable, uses $s$ bits of memory and computes $PartialMOD_n^k(X)$.
\end{lemma} 
\Beginproof
Let us assume that we can construct such a deterministic online algorithm $A_s$. Therefore we can construct an automaton $T$, with $2^s$ states that emulates the same algorithm and computes  $PartialMOD_n^k(X)$. But it is impossible, due to \cite{AY12,agky14,agky16}.
\Endproof
\begin{lemma}\label{lm:pp-random-s}
Let integer $s,k$ be such that $s<k=o(\log(n))$, where $n$ is the length of input. Then there is no randomized algorithm that reads an input variable by variable, uses $s$ bits of memory and computes $PartialMOD_n^k(X)$ with probability of error less than $0.5$.
\end{lemma} 
\Beginproof
We can prove the claim using the same technique as in Lemma \ref{lm:pp-deterministic-s}, because of the same lower bound for probabilistic automata \cite{AY12,agky14,agky16}.
\Endproof

Now we can discuss deterministic online algorithms for $(n,k,r,w)$-PNH Problem.

\begin{theorem}\label{th:phn-deterministic-s}
Let integer $s,k$ be such that $s<k=o(\log(n))$, where $n$ is the length of input. Any deterministic online algorithm $A_s$ computing  $(n,k,w,r)$-PNH Problem is $(w/r)$-competitive.
\end{theorem}
\Beginproof
Let us assume that we have such an algorithm $A_s$. Then we suggest the input $I=(x_1,\dots,x_n)$ such that $A_s$ returns the wrong answer on all requests of  guardians.  Let $m_1=m_2=m_3=m=3\cdot 2^k$.
 
The first guardian answers $y_i$. Due to Lemma \ref{lm:pp-deterministic-s} we can choose input $X_1\in\{0,1\}^{m}$ such that  $A_s$ cannot compute $PartialMOD_m^k(X_1)$. It means that we can choose $X^1$ such that $y_2=v_1\oplus y_1$, for $ v_1=(\#_1(X_1)/2^{k}))$ $mod$ $2$. By the same reason we can pick $X^2$ such that $y_3=v_2\oplus y_2$, for $ v_2=(\#_1(X_2)/2^{k}))$ $mod$ $2$.
Let us choose the input $X^3$ such that $v_3\oplus y_3=1$, for $v_3=(\#_1(X_3)/2^{k}))$ $mod$ $2$.  
  Note that we guarantee that $\#_1(X_i)/2^{k}$ is an integer, for $i\in\{1,2,3\}$.

Therefore, we have:
$z_3=(\#_1(X^3)/2^{k})\mbox{ }mod\mbox{ }2\neq y_3,$
$z_2=(\#_1(X_2,X_3)/2^{k})\mbox{ }mod\mbox{ }2=v_2\oplus v_3\neq y_2,$
$z_1=(\#_1(X_1,X_2,X_3)/2^{k})\mbox{ }mod\mbox{ }2=v_1\oplus v_2\oplus v_3\neq y_1.$
We got a contradiction for input $(2,X^1,2,X^2,2,X^3)$.
Hence the cost of output is $w$ and the competitive ration is $w/r$.  
\Endproof

\begin{theorem}\label{th:pnh-random-s}
Let integer $s,k$ be such that $s<k=o(\log(n))$, where $n$ is the length of input. Any randomize online algorithm $R_s$ computing  $(n,k,w,r)$-PNH Problem is $(r+7w)/(8r)$-competitive.
\end{theorem}
\Beginproof
By the same way as in the previous Theorem we can show that for any algorithm $R_s$ we can suggest the input such that it cannot say anything better than just guessing answers with probabilities $0.5$. Therefore expected cost is $(r+7w)/8$ and expected competitive ratio is $(r+7w)/(8r)$.
\Endproof

Note that for any deterministic online algorithm without memory restriction, we can construct input such that at least two of three guardians return wrong answers.

\begin{theorem}\label{th:deterministic-pnh-s}
Suppose a deterministic online algorithm $A$ computes $(n,k,w,r)$-PNH; then $A$ is $(w/r)$-competitive.
\end{theorem} 
\Beginproof  Let the algorithm $A$ receives the input $I=(x_1,\dots,x_{n})=(2,X_1,2,X_2,2,X_3)$, such that $X_1,X_2,X_3\in\{0,1\}^{m}$, for $m=3\cdot 2^k$. Let $X_1,X_2$ be such that $PartialMOD_m^k(X_1)=PartialMOD_m^k(X_2)=0$.

Then $A$ receives part $(2,X_1,2,X_2,2)$ of the input  and returns $y_1,y_2,y_3$. Let $b=1$, if $y_1+y_2+y_3\geq 2$; and $b=0$, otherwise.  Then we choose $X_3$ such that $PartialMOD_m^k(X_3)\neq b$. In that case $z_1=z_2=z_3= PartialMOD_m^k(X_3)\neq b$. Therefore at least two of three guardians return wrong answers. Therefore $cost(I,A(I))=w$ and $A$ is $(w/r)$-competitive. 3 
\Endproof

It is easy to see that $(r+7w)/(8r)>(w+r)/(2r)$ and $w/r>(w+r)/(2r)$ due to $r<w$. Therefore, a pure quantum algorithm is better than any deterministic or randomize online algorithm that uses less than $k$ bits of memory. And the same pure quantum online algorithm is better than any deterministic online algorithm without memory restriction. 

\subsection{Polylogarithmic Space Complexity Separation between Quantum and Deterministic Online Algorithms with Polylogarithmic Memory}

 Note, that above results show separation for sublogarithmic memory. For polylogarithmic memory case, we present separation between quantum and deterministic models in Theorems \ref{th:phen-quantum-s} and \ref{th:phen-deterministic-s}. Also, same results are right in logarithmic memory case.

Let us consider modification of $(n,k,r,w)$-PNH problem called $(n,r,w)$-Parity Number of Equality Hats or $(n,r,w)$-PNEH. It is the same problem, but we use $EQ_m(X)$ function instead of $PartialMOD_m^k$. Boolean function $EQ_m:\{0,1\}^n\to\{0,1\}$ is such that $EQ(x_1,\dots x_{\lfloor m/2\rfloor}, x_{\lfloor m/2\rfloor+1}, \dots x_{m})=1$, if $(x_1,\dots x_{\lfloor m/2\rfloor})=(x_{\lfloor m/2\rfloor+1}, \dots x_{m})$, and $0$ otherwise. So $z_j(I)=\bigoplus_{i=j}^3 EQ_{m_i}(X_i)$. We suppose that $m>1, r<w$.

Let us construct a pure quantum online algorithm that uses $O(\log  n)$ and solves (n,r,w)$-PNEH$.

{\bf Algorithm 2. (Quantum Algorithm for $(n,r,w)$-PNEH)} The pure quantum algorithm $Q=Q_{O(\log {n})}$ uses $O(\log {n})$ qubits.

{\em Step 1.} The algorithm emulates guessing for $z_1(I)$. $Q$ initializes the qubit $|\psi\rangle=|\frac{1}{\sqrt{2}}|0\rangle+\frac{1}{\sqrt{2}}|1\rangle$. And it measures the qubit before reading any input variables. It gets $|0\rangle$ or $|1\rangle$ with equal probability. The result of measurement is $y_1$. 

\textit{Step 2.}
The algorithm takes $X_1$ and obtains a value of the function $v_1=EQ(X_1)$ using  \textit{quantum fingerprinting method} presented in \cite{av2009,av2008,akv2008}, \cite{an2008,an2009,af98}. This method allows to compute the function $EQ_{m_1}$ with probability of error by any fixed constant $\varepsilon>0$. The method uses $O(\log {m_1})$ qubits. For $0$-instances, probability of error is $\varepsilon$. And for $1$-instances probability of error is $0$. 

\textit{Step 3.} If $Q$ meets $2$ then it takes  a value $v_1$. Let the value be in qubit $|\phi\rangle$. Then the algorithm applies $CNOT$ gate for $|\phi\rangle|\psi\rangle$. After that $|\psi\rangle=|y_1\oplus v_1\rangle$. Then $Q$ measure $|\psi\rangle$ and returns $y_2$.

{\em Step 4.} The step is similar to Step 2, but algorithm reads $X_2$.

{\em Step 5.} The step is similar to Step 3, but algorithm outputs $y_3$.

{\em Step 6.} The algorithm reads and skips the last part of the input. $Q$ does not need these variable, because it guesses $y_1$ and using this value we already can obtain $y_2$ and $y_3$ without $X_3$.

Let us compute an expected cost of pairs $(I,Q(I))$. For this we construct a Table \ref{tbl:pneh} of probabilities and costs  for all possible values $(z_1,z_2,z_3)$ in columns and all possible answers $(y_1,y_2,y_3)$ in rows.



\begin{table}[]
\centering
\caption{Probabilities and costs}
\label{tbl:pneh}
\begin{tabular}{@{}|
>{\columncolor[HTML]{EFEFEF}}l |l|l|l|l|l|l|l|l|@{}}
\toprule

& \cellcolor[HTML]{EFEFEF}\textbf{000} 
& \cellcolor[HTML]{EFEFEF}\textbf{001} 
& \cellcolor[HTML]{EFEFEF}\textbf{010} 
& \cellcolor[HTML]{EFEFEF}\textbf{011} 
& \cellcolor[HTML]{EFEFEF}\textbf{100} 
& \cellcolor[HTML]{EFEFEF}\textbf{101} 
& \cellcolor[HTML]{EFEFEF}\textbf{110} 
& \cellcolor[HTML]{EFEFEF}\textbf{111} \\ \midrule

\textbf{000}  &$ P=\frac{(1-\varepsilon)^2}{2}$&$ P=\frac{(1-\varepsilon)^2}{2}$ & $P=0$ & $P=0$ & $P=0$ & $P=0$ & $P=0$ & $P=0$\\& $c=r$ & $c=w$  
&   &        &         &      &       &    \\ \midrule
\textbf{001}  & $P=\frac{(1-\varepsilon)\varepsilon}{2}$ & $P=\frac{(1-\varepsilon)\varepsilon}{2}$ & $P=\frac{(1-\varepsilon)}{2}$ & $P=\frac{(1-\varepsilon)}{2}$& $P=0$ & $P=0$ & $P=0$ & $P=0$ 
\\& $c=w$     &  $c=w$     &  $c=w$    & $c=r$       &         &      &       &    \\ \midrule

\textbf{010}  & $P=\frac{\varepsilon ^2}{2}$& $P=\frac{\varepsilon ^2}{2}$& $P=\frac{\varepsilon}{2}$& $P=\frac{\varepsilon}{2} $ & $P=\frac{\varepsilon}{2} $ & $P=\frac{\varepsilon}{2} $ & $P=\frac{1}{2} $& $P=\frac{1}{2} $\\& $c=w$    &  $c=w$       &  $c=w$      &    $c=w$      &      $c=w$     &  $c=w$      &  $c=r$       &  $c=w$    \\ \midrule

\textbf{011}  & $P=\frac{(1-\varepsilon)\varepsilon}{2}$& $P=\frac{(1-\varepsilon)\varepsilon}{2}$ &
$P=0$ & $P=0$ & $P=\frac{(1-\varepsilon)}{2}$& $P=\frac{(1-\varepsilon)}{2}$&$P=0$ & $P=0$ \\
& $c=w$     &   $c=w$     &      &        &   $c=w$       &     $c=r$   &       &    \\ \midrule
\textbf{100}  &$P=\frac{(1-\varepsilon)\varepsilon}{2}$&$P=\frac{(1-\varepsilon)\varepsilon}{2}$&$P=0$ & $P=0$ & $P=\frac{(1-\varepsilon)}{2}$& $P=\frac{(1-\varepsilon)}{2}$&$P=0$ & $P=0$\\& $c=w$    &   $c=w$     &      &        &   $c=r$       &     $c=w$   &       &    \\ \midrule

\textbf{101}  & $P=\frac{\varepsilon ^2}{2}$& $P=\frac{\varepsilon ^2}{2}$& $P=\frac{\varepsilon}{2}$& $P=\frac{\varepsilon}{2} $ & $P=\frac{\varepsilon}{2} $ & $P=\frac{\varepsilon}{2} $ & $P=\frac{1}{2} $& $P=\frac{1}{2} $\\ &$c=w$    &   $c=w$        &  $c=w$      &    $c=w$      &      $c=w$     &  $c=w$      &  $c=w$       &  $c=r$     \\ \midrule
\textbf{110}  & $P=\frac{(1-\varepsilon)\varepsilon}{2}$ & $P=\frac{(1-\varepsilon)\varepsilon}{2}$& $P=\frac{(1-\varepsilon)}{2}$& $P=\frac{(1-\varepsilon)}{2}$&$P=0$ & $P=0$&$P=0$ & $P=0$\\& $c=w$     &  $c=w$      &   $c=r$    &    $c=w$     &         &      &       &    \\ \midrule
\textbf{111}  & $P=\frac{(1-\varepsilon)}{2}^2$ & $P=\frac{(1-\varepsilon)}{2}^2$&$P=0$ & $P=0$&$P=0$ & $P=0$&$P=0$ & $P=0$\\& $c=w$     &    $c=r$      &      &        &         &      &       &    
\\ \bottomrule
\end{tabular}
\end{table}

Let us compute expected cost for each kind of inputs $I_{z_1z_2z_3}$:

$cost(Q(I_{000}))=cost(Q(I_{001}))=r(1-\varepsilon)^2/2 + w\big((1-\varepsilon^2)/2+\varepsilon\big)$

$cost(Q(I_{010}))=cost(Q(I_{011}))=cost(Q(I_{100}))=cost(Q(I_{101}))=r(1-\varepsilon)/2+w(1+\varepsilon)/2$

$cost(Q(I_{110}))=cost(Q(I_{111}))=r/2+w/2$

So, expected ratio is $(r(1-\varepsilon)^2/2 + w(\frac{1-\varepsilon^2}{2}+\varepsilon))/r$.
As a result we have the following theorem:

\begin{theorem}\label{th:phen-quantum-s} 
There is $(r(1-\varepsilon)^2/2 + w(\frac{1-\varepsilon^2}{2}+\varepsilon))/r$-competitive in expectation pure quantum algorithms for $(n,r,w)$-PNEH Problem $Q_{O(\log {n})}$ with $O(\log {n})$ qubits of memory. \end{theorem}

At the same time, if a deterministic online algorithm for $(n,r,w)$-PNEH Problem uses a polylogarithmic number of bits; then it is $(w/r)$-competitive. To show this claim, let us discuss a required property of $EQ_m$ function.

\begin{lemma}\label{lm:eq-deterministic-s}
There is no deterministic algorithm that reads an input variable by variable, uses $s=o(m)$ bits of memory and computes $EQ_m(X)$.
\end{lemma} 
\Beginproof
Let us assume that we can construct such a deterministic online algorithm $A_s$. Therefore we can construct an automaton $T$, with $2^s=2^{o(n)}$ states that emulates the same algorithm and computes  $EQ_m(X)$. But it is easy to see that $T$ requires at least $2^{O(n)}$ states.
\Endproof

Now we can discuss deterministic online algorithms for $(n,r,w)$-PNEH Problem.

\begin{theorem}\label{th:phen-deterministic-s}
Let integer $s$ be such that $s=o(n)$, where $n$ is the length of input. Any deterministic online algorithm $A_s$ computing  $(n,w,r)$-PNEH Problem is $(w/r)$-competitive.
\end{theorem}
\Beginproof
Using Lemma \ref{lm:eq-deterministic-s}, we can prove the theorem by the same way as in proof of Theorem \ref{th:phn-deterministic-s}.
\Endproof

Note that for any deterministic online algorithm without memory restriction, we can construct input such that at least two of three guardians return wrong answers.

\begin{theorem}\label{th:deterministic-pneh-s}
Suppose a deterministic online algorithm $A$ computes $(n,k,w,r)$-PNEH; then $A$ is $(w/r)$-competitive.
\end{theorem} 
\Beginproof
We can prove the theorem by the same way as in proof of Theorem \ref{th:deterministic-pnh-s}.
\Endproof

It is easy to see that $(r(1-\varepsilon)^2/2 + w(\frac{1-\varepsilon^2}{2}+\varepsilon))/r<W/r$ due to $r<w$. Therefore, the pure quantum algorithm is better than any deterministic online algorithm without memory restriction. 
\subsection{On an Emulation of a Probabilistic Online Algorithm by a Quantum One}

Let us show that a quantum model has at least the same power as a classical model. We use a simple technique, but it is important to show this result, because as we discussed in the introduction, for some models, quantum and classical cases are incomparable.
 
  \begin{theorem}\label{th:emulation} Let an online optimization problem ${\tt S}$ is solved by $c$-competitive randomized online algorithm $A$ using $s$ bits of memory, then there is a $c$-competitive quantum online algorithm $Q$ that computes this problem using at most $s$ bits of classical memory and one qubit. (See Appendix \ref{apx:emulation})
  \end{theorem} 
 
Pure quantum algorithms also can emulate random algorithms.

  \begin{theorem}\label{th:emulation-pure} Let an online optimization problem ${\tt S}$ is solved by $c$-competitive randomized online algorithm $A$ using $s$ bits of memory, then there is a $c$-competitive pure quantum online algorithm $Q$ that computes ${\tt S}$ using at most $s+1$ qubits. 
  \end{theorem} 
  \Beginproof We can use the same technique as in the previous theorem, but we will store a state of classical memory in the quantum bits.\Endproof

\section{Advice complexity of Quantum online algorithms}\label{sec:advice}
\subsection{Results on Model with Private Qubits}\label{sec:privat}
Firstly, let us show that situation with advice complexity is similar to model with restricted memory, and quantum model with advice  also has at least the same power as a classical one. All proofs of theorems from this section are presented in Appendix \ref{apx:a1}. 
  \begin{theorem}\label{th:emulation-advice} Let an online optimization problem ${\tt S}$ is solved by $c$-competitive randomized online algorithm $A$ with $b$ advice bits, then there is a $c$-competitive quantum online algorithm $Q$ that computes ${\tt S}$ using at most $b$ advice qubits. \end{theorem} 

Let us apply the above theorem to the $(n,k,r,w)$-PNH problem.
We have shown in Theorem \ref{th:deterministic-pnh-s}, which for any deterministic online algorithm, we can construct an input such that at least two of three guardians return wrong results. At the same time, one advice bit is enough to construct optimal an online solution.  

\begin{theorem}\label{th:deterministic-pnh}
There is optimal deterministic online algorithm $B$  with $1$ advice bit for $(n,k,r,w)$-PNH problem. 
\end{theorem}

We will formulate two quantum online algorithms for the problem. The first one emulates a random algorithm, the second one is an optimal pure quantum online algorithm with advice, and it is based on ideas from Theorem \ref{th:phn-quantum-s}.

\begin{theorem}\label{th:random-quantum-pnh}
There are $(w+r)/(2r)$-competitive randomized online algorithm $R$ and $(w+r)/(2r)$-competitive quantum online algorithm $Q$  for $(n,k,r,w)$-PNH Problem.
\end{theorem}

\begin{theorem}\label{th:pure-quantum-pnh}
There is an optimal pure quantum algorithm $Q$ with single advice qubit that uses a single qubit of memory for solving $(n,k,r,w)$-PNH Problem. 
\end{theorem}

\subsection{Results on Model with Shared Entangled Qubits}\label{sec:shared}
For any classical online algorithm with advice, we can construct a quantum algorithm with shared qubits and two times less advice qubits. The idea is based on paper \cite{bw92}, where authors use EPR-pairs. It allows to send 2 bits of classical information using one qubit.
 
  \begin{theorem}\label{th:advice-epr}Let $P$ be an online minimization problem, and $A$ be $c_A$-competitive deterministic online algorithm with $b_A$ advice bits, $R$ be $c_B$-competitive randomized online algorithm with $b_B$ advice bits. Then, there are $c_A$-competitive quantum online algorithm with shared qubits and $\lceil b_A/2 \rceil$ advice qubits, and $c_B$-competitive quantum online algorithm with shared qubits and $\lceil b_B/2 \rceil$ advice qubits for the same problem. (See Appendix \ref{apx:advice-epr})\end{theorem}  

{\bf The Paging problem.}
Let us apply this result to the well-known Paging problem \cite{dkp2009}.  We describe it in a very simplified way; a practical view of the problem can be found in the standard literature, for example, in \cite{k2016}.  

  Consider a two-level memory system that consists of a small fast memory and a large slow memory. Here, each request specifies a page in the memory system. A request is served if the corresponding page is in the fast memory. If a requested page is not in the fast memory, a page fault occurs. Then some page must be moved from the fast memory to the slow memory so that the requested page can be loaded into the vacated location. A paging algorithms specifies which page to evict on a fault. The cost is the total number of evicted pages.  
  The large slow memory has $N$ pages. The fast memory (cache) has $k < N$ pages.

\begin{theorem}[\cite{dkp2009}] There is an optimal deterministic online algorithm $A$ for Paging Problem, which uses $n$ bits of advice.\end{theorem}
 
  Hence, for the quantum case, we get the following result due to Theorem \ref{th:advice-epr}.  
\begin{theorem} There is an optimal quantum online algorithm $A$ with shared qubits for Paging Problem, which uses $n/2$ qubits of advice. \end{theorem}

{\bf Acknowledgements. }
Partially supported by ERC Advanced Grant MQC. The work is performed according to the Russian Government Program of Competitive Growth of Kazan Federal University

\bibliographystyle{alpha}
\bibliography{tcs}

\newcommand{\etalchar}[1]{$^{#1}$}
\begin{thebibliography}{GKK{\etalchar{+}}07}

\bibitem[AAKK17]{aakk2017}
Farid Ablayev, Andris Ambainis, Kamil Khadiev, and Aliya Khadieva.
\newblock Lower bounds and hierarchies for quantum automata communication
  protocols and quantum ordered binary decision diagrams with repeated test.
\newblock 2017.

\bibitem[AF98]{af98}
Andris Ambainis and R\={u}si\c{n}\v{s} Freivalds.
\newblock 1-way quantum finite automata: strengths, weaknesses and
  generalizations.
\newblock In {\em FOCS'98: Proceedings of the 39th Annual Symposium on
  Foundations of Computer Science}, pages 332--341, 1998.
\newblock (http://arxiv.org/abs/quant-ph/9802062).

\bibitem[AGK{\etalchar{+}}05]{agkmp2005}
Farid Ablayev, Aida Gainutdinova, Marek Karpinski, Cristopher Moore, and
  Christopher Pollett.
\newblock On the computational power of probabilistic and quantum branching
  program.
\newblock {\em Information and Computation}, 203(2):145--162, 2005.

\bibitem[AGKY14]{agky14}
Farid Ablayev, Aida Gainutdinova, Kamil Khadiev, and Abuzer Yakaryılmaz.
\newblock Very narrow quantum obdds and width hierarchies for classical obdds.
\newblock In {\em Descriptional Complexity of Formal Systems}, volume 8614 of
  {\em Lecture Notes in Computer Science}, pages 53--64. Springer, 2014.

\bibitem[AGKY16]{agky16}
F.~Ablayev, A.~Gainutdinova, K.~Khadiev, and A.~Yakary{\i}lmaz.
\newblock Very narrow quantum obdds and width hierarchies for classical obdds.
\newblock {\em Lobachevskii Journal of Mathematics}, 37(6):670--682, 2016.

\bibitem[AKV10]{akv2008}
Farid Ablayev, Airat Khasianov, and Alexander Vasiliev.
\newblock On complexity of quantum branching programs computing equality-like
  boolean functions.
\newblock {\em ECCC}, 2010.

\bibitem[Alb96]{a1996}
Susanne Albers.
\newblock {\em BRICS, Mini-Course on Competitive Online Algorithms}.
\newblock Aarhus University, 1996.

\bibitem[AN08]{an2008}
Andris Ambainis and Nikolajs Nahimovs.
\newblock Improved constructions of quantum automata.
\newblock In {\em TQC}, pages 47--56. Springer, 2008.

\bibitem[AN09]{an2009}
Andris Ambainis and Nikolajs Nahimovs.
\newblock Improved constructions of quantum automata.
\newblock {\em Theoretical Computer Science}, 410(20):1916--1922, 2009.

\bibitem[AV08]{av2008}
Farid Ablayev and Alexander Vasiliev.
\newblock On the computation of boolean functions by quantum branching programs
  via fingerprinting.
\newblock In {\em Electronic Colloquium on Computational Complexity (ECCC)},
  volume~15, 2008.

\bibitem[AV09]{av2009}
Farid~Mansurovich Ablayev and AV~Vasilyev.
\newblock On quantum realisation of boolean functions by the fingerprinting
  technique.
\newblock {\em Discrete Mathematics and Applications}, 19(6):555--572, 2009.

\bibitem[AY12]{AY12}
Andris Ambainis and Abuzer Yakary{\i}lmaz.
\newblock Superiority of exact quantum automata for promise problems.
\newblock {\em Information Processing Letters}, 112(7):289--291, 2012.

\bibitem[AY15]{AY15}
Andris Ambainis and Abuzer Yakary{\i}lmaz.
\newblock Automata and quantum computing.
\newblock Technical Report 1507.01988, arXiv, 2015.

\bibitem[BEY05]{be2005}
Allan Borodin and Ran El-Yaniv.
\newblock {\em Online computation and competitive analysis}.
\newblock cambridge university press, 2005.

\bibitem[BFK{\etalchar{+}}16]{bfklm2016}
Joan Boyar, Lene~M Favrholdt, Christian Kudahl, Kim~S Larsen, and Jesper~W
  Mikkelsen.
\newblock Online algorithms with advice: A survey.
\newblock {\em Acm Sigact News}, 47(3):93--129, 2016.

\bibitem[BFK{\etalchar{+}}17]{bfklm2017}
Joan Boyar, Lene~M Favrholdt, Christian Kudahl, Kim~S Larsen, and Jesper~W
  Mikkelsen.
\newblock Online algorithms with advice: A survey.
\newblock {\em ACM Computing Surveys (CSUR)}, 50(2):19, 2017.

\bibitem[BIL09]{bil2009}
Joan Boyar, Sandy Irani, and Kim~S Larsen.
\newblock A comparison of performance measures for online algorithms.
\newblock In {\em Workshop on Algorithms and Data Structures}, pages 119--130.
  Springer, 2009.

\bibitem[BIL15]{bil2015}
Joan Boyar, Sandy Irani, and Kim~S Larsen.
\newblock A comparison of performance measures for online algorithms.
\newblock {\em Algorithmica}, 72(4):969--994, 2015.

\bibitem[BK09]{bk2009}
Luca Becchetti and Elias Koutsoupias.
\newblock Competitive analysis of aggregate max in windowed streaming.
\newblock In {\em Automata, Languages and Programming: 36th International
  Colloquium, ICALP 2009, Rhodes, Greece, July 5-12, 2009, Proceedings, Part
  I}, pages 156--170, Berlin, Heidelberg, 2009. Springer Berlin Heidelberg.

\bibitem[BLM15]{blm2015}
Joan Boyar, Kim~S Larsen, and Abyayananda Maiti.
\newblock The frequent items problem in online streaming under various
  performance measures.
\newblock {\em International Journal of Foundations of Computer Science},
  26(4):413--439, 2015.

\bibitem[BW92]{bw92}
Charles~H Bennett and Stephen~J Wiesner.
\newblock Communication via one-and two-particle operators on
  einstein-podolsky-rosen states.
\newblock {\em Physical review letters}, 69(20):2881, 1992.

\bibitem[DKP09]{dkp2009}
Stefan Dobrev, Rastislav Kr{\'a}lovi{\v{c}}, and Dana Pardubsk{\'a}.
\newblock Measuring the problem-relevant information in input.
\newblock {\em RAIRO-Theoretical Informatics and Applications}, 43(3):585--613,
  2009.

\bibitem[DLO05]{dl2005}
Reza Dorrigiv and Alejandro L{\'o}pez-Ortiz.
\newblock A survey of performance measures for on-line algorithms.
\newblock {\em SIGACT News}, 36(3):67--81, 2005.

\bibitem[Gai15]{g15}
A.~F. Gainutdinova.
\newblock Comparative complexity of quantum and classical obdds for total and
  partial functions.
\newblock {\em Russian Mathematics}, 59(11):26--35, 2015.

\bibitem[GK15]{gk2015}
Yiannis Giannakopoulos and Elias Koutsoupias.
\newblock Competitive analysis of maintaining frequent items of a stream.
\newblock {\em Theoretical Computer Science}, 562:23--32, 2015.

\bibitem[GKK{\etalchar{+}}07]{gkkrw2007}
Dmitry Gavinsky, Julia Kempe, Iordanis Kerenidis, Ran Raz, and Ronald De~Wolf.
\newblock Exponential separations for one-way quantum communication complexity,
  with applications to cryptography.
\newblock In {\em Proceedings of the thirty-ninth annual ACM symposium on
  Theory of computing}, pages 516--525. ACM, 2007.

\bibitem[Hro05]{h2005}
JI~Hromkovic.
\newblock Z{\'a}mecnikov{\'a}. design and analysis of randomized algorithms:
  Introduction to design paradigms, 2005.

\bibitem[IKP{\etalchar{+}}17]{ikpvy2017}
Rishat Ibrahimov, Kamil Khadiev, Kri\v{s}j\={a}nis Pr\={u}sis, Jevgenijs
  Vihrovs, and Abuzer Yakary{\i}lmaz.
\newblock Zero-error affine, unitary, and probabilistic obdds.
\newblock {\em arXiv preprint arXiv:1703.07184}, 2017.

\bibitem[KK17]{kk2017}
Kamil Khadiev and Aliya Khadieva.
\newblock {\em Reordering Method and Hierarchies for Quantum and Classical
  Ordered Binary Decision Diagrams}, pages 162--175.
\newblock Springer International Publishing, Cham, 2017.

\bibitem[Kla00]{k2000}
Hartmut Klauck.
\newblock Quantum communication complexity.
\newblock In {\em In Proc. Intl. Colloquium on Automata, Languages, and
  Programming (ICALP}. Citeseer, 2000.
\newblock arXiv preprint quant-ph/0005032.

\bibitem[KMRS86]{kmrs86}
Anna~R Karlin, Mark~S Manasse, Larry Rudolph, and Daniel~D Sleator.
\newblock Competitive snoopy caching.
\newblock In {\em Foundations of Computer Science, 1986., 27th Annual Symposium
  on}, pages 244--254. IEEE, 1986.

\bibitem[Kom16]{k2016}
Dennis Komm.
\newblock {\em An Introduction to Online Computation: Determinism,
  Randomization, Advice}.
\newblock Springer, 2016.

\bibitem[KT06]{kt2006}
Jon Kleinberg and Eva Tardos.
\newblock {\em Algorithm design}.
\newblock Pearson Education India, 2006.

\bibitem[LG06]{l2006}
Fran{\c{c}}ois Le~Gall.
\newblock Exponential separation of quantum and classical online space
  complexity.
\newblock In {\em Proceedings of the eighteenth annual ACM symposium on
  Parallelism in algorithms and architectures}, pages 67--73. ACM, 2006.

\bibitem[MR10]{mr2006}
Rajeev Motwani and Prabhakar Raghavan.
\newblock {\em Randomized algorithms}.
\newblock Chapman \& Hall/CRC, 2010.

\bibitem[SS05]{ss2005}
Martin Sauerhoff and Detlef Sieling.
\newblock Quantum branching programs and space-bounded nonuniform quantum
  complexity.
\newblock {\em Theoretical Computer Science}, 334(1):177--225, 2005.

\bibitem[ST85]{st85}
Daniel~D Sleator and Robert~E Tarjan.
\newblock Amortized efficiency of list update and paging rules.
\newblock {\em Communications of the ACM}, 28(2):202--208, 1985.

\bibitem[Weg00]{Weg00}
Ingo Wegener.
\newblock {\em Branching Programs and Binary Decision Diagrams: Theory and
  Applications}.
\newblock SIAM, 2000.

\end{thebibliography}

\newpage
\appendix
\section{The proof of Theorem \ref{th:emulation}}\label{apx:emulation}
 Let us construct the quantum algorithm $Q$ which emulates the algorithm $A$. Let $A$ be a randomized online algorithm which uses $d$ random bits from random tape. Note that a deterministic online algorithm is the partial case of a randomized algorithm with $d=0$.
  Before reading the input, $Q$ initializes a qubit $|\psi\rangle=|0\rangle$. In each case, when the algorithm $Q$ should emulate reading of a random bit from random tape, it applies to $|\psi\rangle$ the Hadamar transformation $H$. Here
  $H=\frac{1}{\sqrt{2}}\begin{pmatrix} 1, & 1 \\ 1,& -1\end{pmatrix}$.
  After that $Q$ measures the qubit and gets $1$ or $0$ with probability $0.5$, this action emulates the uniform distributed random bit.  
  By construction, $Q$ returns the same results with the same probability as $A$.
\section{Proofs of Theorems \ref{th:emulation-advice}, \ref{th:deterministic-pnh}, \ref{th:random-quantum-pnh}, \ref{th:pure-quantum-pnh}}\label{apx:a1}
  {\bf Theorem \ref{th:emulation-advice}.} {\em Let an online optimization problem ${\tt S}$ is solved by $c$-competitive randomized online algorithm $A$ with $b$ advice bits, then there is a $c$-competitive quantum online algorithm $Q$ that computes ${\tt S}$ using at most $b$ advice qubits. }
  
  \Beginproof Let us construct the quantum algorithm $Q$ which emulates the algorithm $A$. Let $A$ be a randomized online algorithm which uses $d$ random bits from random tape. Note that deterministic online algorithm is the partial case of a randomized algorithm with $d=0$.
The {\em Adviser} sends $b$ qubits with the same information as for $A$, using pure states of qubits. When $Q$ gets advice bits, then it measures them immediately and gets classical advice bits.
After that $Q$ does the same actions as $A$.
  Let us describe the emulation of the random process. Before reading the input, $Q$ initializes qubit $|\psi\rangle$ in $|0\rangle$ state. In each case, when Algorithm $Q$ is required to emulate reading of a random bit from random tape, it applies the Hadamard transformation $H$ to $|\psi\rangle$  
  After that $Q$ measures the qubit and gets $1$ or $0$ with probability $0.5$, this action emulates the uniform distributed random bit.  
  By construction, $Q$ returns result with the same probability as $A$.\Endproof
   
{\bf Theorem \ref{th:deterministic-pnh}.} {\em 
There is an optimal deterministic online algorithm $B$  with $1$ advice bit for $(n,k,r,w)$-PNH problem. 
}

\Beginproof  {\em Adviser} sends parity of numbers of $1$s over $2^k$. Let the value of this bit be $p$. Algorithm $B$ computes the number of $1$s, when it gets full $2^k$ ones, then it inverts $p$. The algorithm returns $p$ on each request. It is easy to see that $B$ always returns right answers. Thus it is optimal.
\Endproof

{\bf Theorem \ref{th:random-quantum-pnh}.} {\em 
There are $(w+r)/(2r)$-competitive randomized online algorithm $R$ and $(w+r)/(2r)$-competitive quantum online algorithm $Q$  for $(n,k,r,w)$-PNH Problem.
}

\Beginproof
Firstly, algorithm $R$ guesses $z_1$ with equal probability. Let value of guess be $p\in\{0,1\}$. Algorithm $R$ computes the number of $1$s, when it gets full $2^k$ ones, then it inverts $p$. The algorithm returns $p$ for each request. It is easy to see that if the guess is right, then $R$ always returns the right answers, and $R$ always returns the wrong answer otherwise. Therefore, competitive ratio is $(0.5 \cdot r+0.5 \cdot w)/r=(r+w)/(2r)$.
  For the quantum case, we apply Theorem \ref{th:emulation} and get the claim of the theorem.
\Endproof

{\bf Theorem \ref{th:pure-quantum-pnh}.} {\em 
There is an optimal pure quantum algorithm $Q$ with single advice qubit that uses single qubit of memory for solving $(n,k,r,w)$-PNH Problem. 
}

\Beginproof
We can construct an algorithm like in the proof of Theorem \ref{th:phn-quantum-s}. The difference is that the {\em Adviser} sends one qubit $|\psi\rangle$ for $z_1$. \Endproof

\section{The Proof of Theorem \ref{th:advice-epr}}\label{apx:advice-epr}
Let us consider a deterministic online algorithm. The proof for the randomized case is the same.
  We interpreted the sending of advice bits as a communication game between the algorithm and the {\em Adviser}. Let the algorithm wants to receive some information from the {\em Adviser}. To perform this feat, the algorithm prepares an EPR-pair (two qubits in a state $\sqrt{1/2}|01\rangle-\sqrt{1/2}|10\rangle$) and sends one qubit of the pair to the {\em Adviser}. After that, the {\em Adviser} applies some unitary operation described in \cite{bw92}, to the qubit to encode information with one of the states. And {\em Adviser} sends the qubit to the algorithm. Then the pair of qubits can be in one of these states (basis):  
\begin{equation}\label{eq:epr-pairs}
\sqrt{1/2}(|00\rangle+|11\rangle),\sqrt{1/2}(|01\rangle+|10\rangle), \sqrt{1/2}(|01\rangle-|10\rangle), \sqrt{1/2}(|00\rangle-|11\rangle)
\end{equation}
 
Each state encodes two bits of information.
  The algorithm can measure qubits jointly in the orthonormal basis (\ref{eq:epr-pairs}), and so reliably learn which operator the {\em Adviser} applied.
For $b_A$ bits the algorithm prepares $\lceil b_A/2 \rceil$ pairs and sends one qubit for each pair. Then the {\em Adviser} applies the described transformation and sends $\lceil b_A/2 \rceil$ qubits back. After that, the algorithm can get $ b_A$ classical bits from received qubits.

\end{document}